\documentclass[twocolumn,superscriptaddress]{revtex4-2}

\usepackage[utf8]{inputenc}
\usepackage{amsmath}
\usepackage{graphicx}
\usepackage{float}
\usepackage{braket}
\usepackage{amssymb}
\usepackage{color}
\usepackage{siunitx}

\usepackage{subcaption}

\newcommand{\ddt}{\frac{\mathrm{d}}{\mathrm{d}t}}
\newcommand{\rfac}{r_\mathrm{f}}
\newcommand{\gfac}{\Gamma_\mathrm{f}}
\newcommand{\ncrit}{n_\mathrm{crit}}
\newcommand{\kay}{\langle k \rangle}
\newcommand{\drfac}{\delta \rfac}

\begin{document}

\begin{abstract}
The facilitation of Rydberg excitations in a gas of atoms 
provides an ideal model system to study 
%excitation spreading and 
epidemic evolution on (dynamic) networks and self organization of complex systems to the critical point of a non-equilibrium phase transition.  
%For percolating networks an absorbing-state phase transition emerges. 
Using Monte-Carlo simulations and a machine learning algorithm we show that the universality class of this phase transition can be tuned. The classes include directed percolation (DP), the most common
class in short-range spreading models, and  mean-field (MF) behavior, but also different types of anomalous directed percolation (ADP), characterized by rare long-range excitation processes. In a frozen gas, ground state atoms that can facilitate each other form a static network, for which we predict DP universality. Atomic motion then turns the network into a dynamic one with long-range (Lévy-flight type) %character of the 
excitations. This
% is caused by atomic motion and 
leads to continuously varying critical exponents corresponding to the ADP universality class, eventually reaching MF behavior.
%depending on the average atom velocity. 
These findings also explain the recently observed critical exponent of Rydberg facilitation in an ultra-cold gas experiment [Helmrich \textit{et al.}, Nature \textbf{577}, 481 (2020)], which was in between DP and MF values. %\daniel{Last 2 sentences removed!}

%For a frozen gas, which we show to be of DP universality. With increasing average velocity the system however crosses through different ADP phases and eventually approaches MF behavior.
\end{abstract}

%%%%%%%%%%%%%%%%%%%%%%%%%%%%%%%%%%%%%%%%%%%%%%%%%%%%%%%%%%%%%%%%%%%%%%%%%%%%%%%%%%%%%
% \title{Rydberg facilitation in a gas of %moving 
% atoms: Anomalous directed-percolation \mfl{(or: Anomalous non-equilibrium critical behaviour)} on a dynamical network 
% }

\title{Anomalous Directed Percolation on a Dynamic Network using Rydberg Facilitation}
%%%%%%%%%%%%%%%%%%%%%%%%%%%%%%%%%%%%%%%%%%%%%%%%%%%%%%%%%%%%%%%%%%%%%%%%%%%%%%%%%%%%%

%%%%%%%%%%%%%%%%%%%%%%%%%%%%%%%%%%%%%%%%%%%%%%%%%%%%%%
%\title{Rydberg Atoms \emoji{call-me-hand}}
%%%%%%%%%%%%%%%%%%%%%%%%%%%%%%%%%%%%%%%%%%%%%%%%%%%%%%

\author{Daniel Brady}
\affiliation{Department of Physics and Research Center OPTIMAS, RPTU Kaiserslautern, D-67663 Kaiserslautern, Germany}
\author{Simon Ohler}
\affiliation{Department of Physics and Research Center OPTIMAS, RPTU Kaiserslautern, D-67663 Kaiserslautern, Germany}
\author{Johannes Otterbach}
\affiliation{Orthonogal Otter UG, Berlin, Germany}
%\author{Winfried Ripken}
%\affiliation{Merantix Momentum, AI Campus Berlin, Germany}
\author{Michael Fleischhauer}
\affiliation{Department of Physics and Research Center OPTIMAS, RPTU Kaiserslautern, D-67663 Kaiserslautern, Germany}

\date{\today}

\maketitle

% \daniel{Word Count: 2520 / 3000 (10.04.24 13:46)}

% \daniel{Word Count: 2838 / 3000 (12.04.24 14:09)}

% \daniel{Word Count: 2788 / 3000 (15.04.24 10:42)}

Non-equilibrium phase transitions \cite{hinrichsen2000non} and the dynamical self organisation of complex systems to the corresponding
critical point  \cite{bak1988self,bak2013nature} are %widespread 
key phenomena believed to be the reason
for the abundance of scale invariance in nature. They are characteristic for a broad range of spreading processes ranging from epidemic dynamics of diseases \cite{bailey1975mathematical,pastor2015epidemic}, earthquakes \cite{sornette1989self}, and forest fires \cite{drossel1992self}, to neural networks \cite{friedman2012universal}, electric circuits, and information spreading  in the internet \cite{adamic2000power}. 
The most relevant non-equilibrium phase transitions are those between an active and an inactive phase (absorbing state) of dynamical activity. 
In contrast to their equilibrium counterpart, %non-equilibrium phase transitions
they are much less understood and the corresponding theoretical models can in general not be solved exactly. However, in analogy to equilibrium statistical mechanics, the behavior near the critical point shows universal features characterized by different non-equilibrium universality classes
\cite{hinrichsen2000non}.

One of the most prominent universality classes of non-equilibrium phase transitions is directed percolation (DP) \cite{hinrichsen2000non}, originally describing the flow of fluids through porous materials. Janssen and Grassberger conjectured that non-equilibrium phase transitions in \textit{any} classical system should belong to the DP universality class if they: (i) exhibit a continuous phase transition between an active and a unique absorbing state, (ii) the transition is characterized by a  positive one-component order parameter, (iii) the dynamical rules involve only short-range interactions, and (iv) the system has no special attributes such as additional symmetries or quenched randomness \cite{janssen1981nonequilibrium, grassberger1981on}. To date no counterexamples to these criteria have been found \cite{hinrichsen2006non}, and DP universality has even been predicted in more general systems, 
% found in systems which violate one or several of the above points, 
e.g. %in systems 
with multiple absorbing states \cite{munoz1996critical, munoz1998phase}.

%%%%%%%%%%%%%%%%%%%%%%%%%%%%%%%%%%%%%%%%%%%%%%%%%%%%%%%%%%%%%%%%%%%%%%%%%%%%%%%%%%
\begin{figure}[H]
  \centering
  \includegraphics[width=\columnwidth]{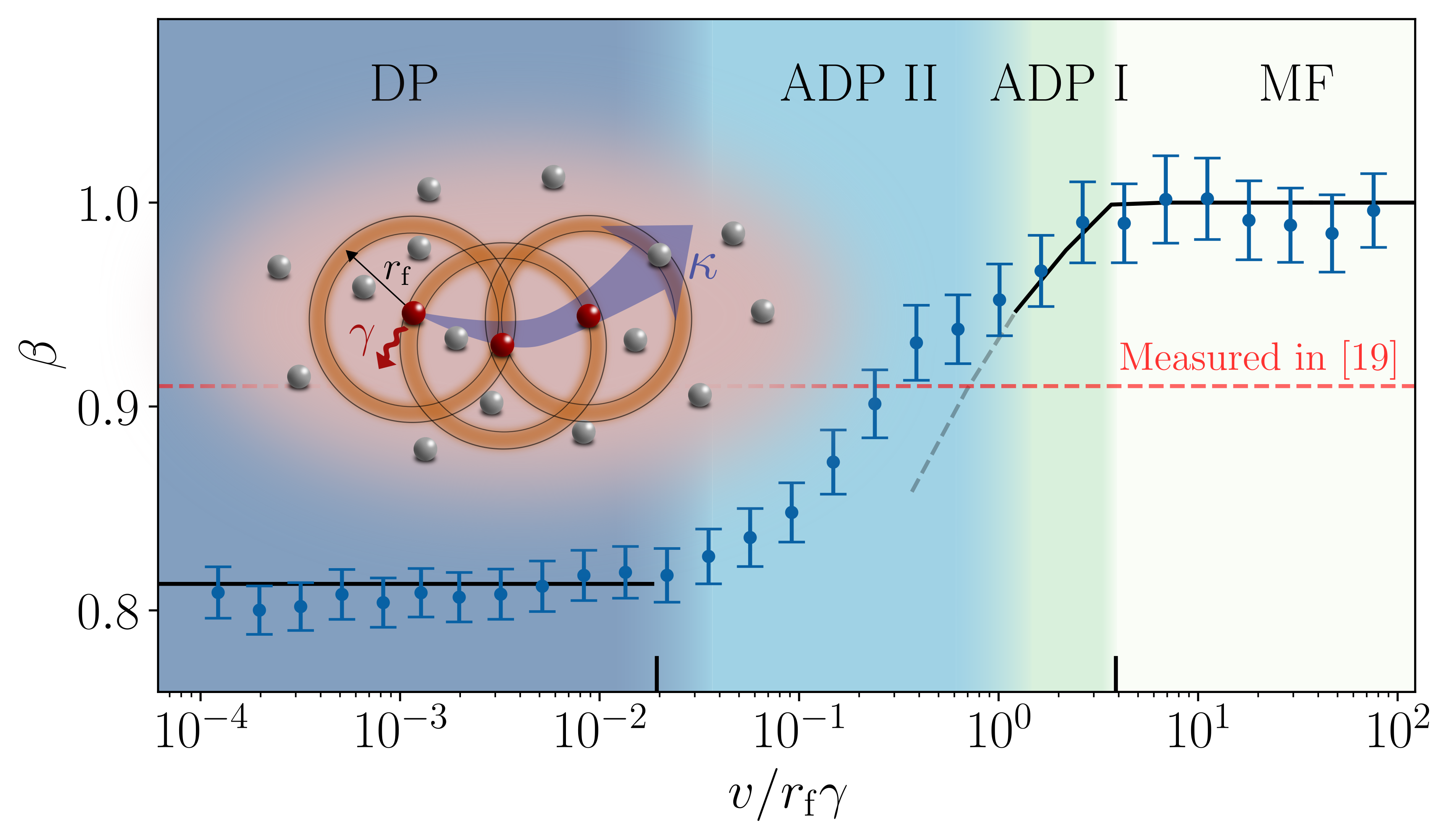}
    \caption{
        Critical exponent $\beta$ of the active density
        % scaling around the critical point 
        as a function of the root mean square % gas 
        atom velocity for a percolating gas with ${\kay = 1.25}$ (see text). The exponent displays two plateaus: 
        % with averaged values 
        ${\beta=0.993(6)}$ for high %gas velocities 
        and ${\beta=0.808(5)}$ for low gas velocities - error given by standard deviation. Mean field %universality (MF) 
        predicts ${\beta_\text{MF} = 1.00}$ and 3D directed percolation %universality predicts 
        ${\beta_\text{DP} \approx 0.813}$. Between these limits the critical exponent changes continuously, characteristic for anomalous directed percolation (ADP). Errors in $\beta$ result from
        %exponent are given by the covariance matrix when 
        fitting (see text). The black line shows a field theoretical approximation of $\beta$ and continues as a grey dashed line in the ADP II phase. The black ticks on the x-axis display relevant velocity scales (see main text). \textit{Inset: }Schematic of spreading dynamics in the gas. Excited Rydberg atoms (red) shift atoms in a spherical shell with radius $\rfac$ and width $\drfac$ into resonance (orange shells). All points use the parameters ${b=0.3}$, ${\Delta / \gamma = 2000}$, with varied ${\Omega / \gamma \in [1, 10]}$ and ${n_0 \rfac^3 \in [20, 30]}$.
    }
\label{fig:beta}
\end{figure}
%%%%%%%%%%%%%%%%%%%%%%%%%%%%%%%%%%%%%%%%%%%%%%%%%%%%%%%%%%%%%%%%%%%%%%%%%%%%%%%%%%

In spite of the apparent generality of the directed percolation  universality class, only few experimental platforms are known 
for which DP behavior has unambiguously been
proven.
% which allow for the systematic study of this behavior. 
In 2007 the first such platform was found in turbulent liquid crystals and a full set of critical exponents in ${d=2+1}$ dimensions was measured \cite{takeuchi2007directed, takeuchi2009experimental}. Since then, interacting many-body systems of Rydberg atoms in the facilitation regime have been suggested as a platform to study absorbing state phase transitions, for which DP universal behavior was predicted on  a lattice \cite{marcuzzi2015non}, and subsequently experimentally observed in a 1D gas \cite{gutierrez2017experimental}.
%Tunable experimental platforms to study non-equilibrium physics and their quantitative modelling are thus of key importance to our understanding of the dynamics of complex systems. 
One important aspect, relevant for the emergence of scale invariance in dynamical processes, %in nature, 
%many real-life processes, 
which
these model systems lack is the effect of atomic losses from the system. In a \textit{number non-conserving regime} the gas density decreases over time which drives the system to its critical point \cite{helmrich2020signatures, ding2020phase}, a phenomenon called self-organized criticality (SOC)
\cite{bak1988self,bak2013nature}. An experiment investigating Rydberg facilitation in a 3D gas, performed in this number non-conserving regime \cite{helmrich2020signatures}, showed signatures of SOC, but 
a deviation from DP universality, %- naively expected from the Janssen-Grassberger conjecture - 
which was attributed to the self-organization process.

Through numerical experiments and analytic considerations we show that this deviation is neither due to the self-organization process \cite{helmrich2020signatures} nor due to heterogeneity \cite{natcom_griffiths}, but results from a violation of the Jansen-Grassberger conditions leading to a dependency of the universality class of the underlying absorbing-state phase transition on the relative velocity of the atoms in the gas. Tuning the parameters which set the reference scale of the atomic velocity, the system can either display DP, mean-field (MF), or anomalous directed percolation (ADP) universality. In Fig.~\ref{fig:beta} numerical predictions of the %universal 
critical scaling exponent of the active density around the critical point can be seen as a function of the root mean square (RMS) thermal velocity of the atoms.
%relative to a characteristic scale, which will be explained later on.

Several relevant real-life spreading processes go beyond the Jansen-Grassberger conjecture and display different universal behavior of the absorbing-state phase transition than DP.
An example is the spread of diseases by flying insects in addition to the spread by direct contact, which violates the condition of
short-range excitations 
\cite{mollison1977spatial,pietronero2012fractals}. Likewise 
spreading processes often take place on dynamical rather than static networks 
\cite{holme2012temporal}. For example in social networks, no individual is in contact with all friends simultaneously all the time. Contacts often change on a time scale comparable with the one of the spreading process itself and thus real contact networks are essentially dynamic \cite{zhao2011entropy}. 

Rydberg facilitation in a gas of cold atoms provides a powerful  platform which opens a way to experimentally study 
transport phenomena on resolvable length and time scales, as well as in various static and dynamic network configurations. It allows for the investigation of different types of non-equilibrium %absorbing-state 
phase transitions beyond DP universality as well as the self organization of complex dynamical systems to the corresponding critical points. 

While so far most experimental work on Rydberg facilitation focused on the dephasing dominated regime, in which the system largely behaves classically, the system can also be studied in the quantum limit by tuning microscopic parameters. Here, as argued in Ref.~\cite{marcuzzi2016absorbing}, universal features different from DP or ADP could emerge when quantum effects become relevant. Furthermore, while facilitation in a frozen gas of atoms (i.e. when the thermal movement is much slower than the spreading dynamics) constitutes a spreading process on a random network \cite{brady2024griffiths}, using advanced trapping techniques (i.e. tweezer arrays) various lattice geometries \cite{Endres2016,Barredo2016,Browaeys2020} can be investigated.

%%%%%%%%%%%%%%%%%%%%%%%%%%%%%%%%%%%%%%%%%%%%%%%%%%%%%%%
\section{Critical scaling at the absorbing-state phase transition }
%%%%%%%%%%%%%%%%%%%%%%%%%%%%%%%%%%%%%%%%%%%%%%%%%%%%%%%

In Rydberg facilitation systems, atoms are continuously driven from the ground to a high-lying Rydberg state by a laser with Rabi frequency $\Omega$ and detuning from resonance $\Delta$, where ${\Delta \gg \Omega}$. As a result of the strong detuning, off-resonant (seed) excitations are strongly suppressed. However, in the presence of a Rydberg atom, other atoms with distance ${r \approx \rfac \equiv \sqrt[6]{\frac{C_6}{\Delta}}}$ are shifted into resonance as a result of the van-der-Waals (vdW) interaction, where $C_6$ is the vdW coefficient.

Consequently, atoms within a spherical shell with volume ${V_s \approx 4\pi \drfac \rfac^2}$ around a Rydberg atom can be \textit{facilitated} (i.e. excited on much faster time scales). Here ${\drfac \approx \frac{\gamma_\perp}{2 \Delta} \rfac}$ is the width of the facilitation shell with $\gamma_\perp$ being the linewidth of the transition. The rate of excitation for atoms within the facilitation shell is given by ${\gfac = 2 \Omega^2 / \gamma_\perp}$. These facilitated excitations can also be interpreted as infection processes, with a global spreading rate given by ${\kappa = \gfac n V_s}$, where $n$ is the gas density. Spontaneous decay of Rydberg atoms back to the ground state then corresponds to recovery with rate $\gamma$. For more details about the microscopic dynamics, we refer the reader to the Methods section.

These systems feature a non-equilibrium phase transition between an absorbing phase, for ${\kappa < \gamma}$, with no excited atoms in the thermodynamic limit, and an active phase, for ${\kappa > \gamma}$, featuring widespread and infinitely long-lived activity. Near the critical driving strength ${\kappa \approx \gamma}$, there is universal behavior characterised by three scaling relations for the Rydberg density
$\rho$, as well as the temporal and spatial correlation lengths, 
$\xi_\parallel$ and $\xi_\perp$ respectively
\begin{subequations}
    \begin{align}
        \label{eq:rho_scaling_dp}
        \rho &\sim (p - p_c)^\beta,
        \\
        \xi_\parallel &\sim |p - p_c|^{-\nu_\parallel},
        \\
        \xi_\perp &\sim |p - p_c|^{-\nu_\perp}.
\end{align}
\end{subequations}
Here ${p - p_c}$ corresponds to the distance of the control parameter from the critical point, and $\beta$, $\nu_\parallel$, $\nu_\perp$ are critical exponents.
Finally, while seed excitations are strongly suppressed, they still occur for finite separation of time scales with the rate ${\tau \sim 1 / \Delta^2}$. 

\begin{figure}[H]
  \centering
  \includegraphics[width=\columnwidth]{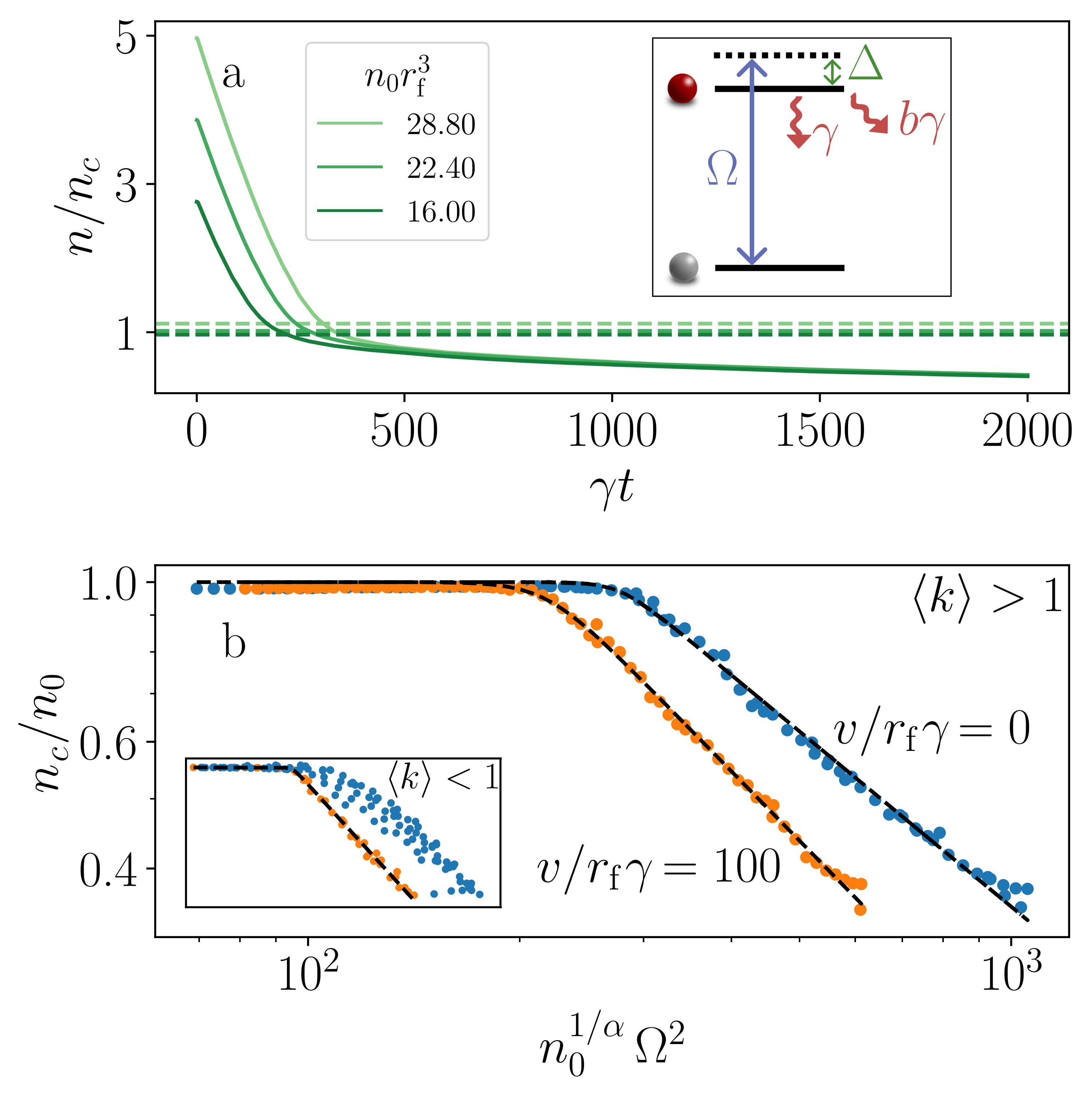}
  \caption{
        (a) Time dynamics of density of ground and Rydberg atoms in the high temperature limit for ${b = 0.3}$, ${\Omega / \gamma = 3.7}$, ${\Delta / \gamma = 1000}$, and ${v / \rfac \gamma = 100}$, with varied $n_0$ showing self-organized criticality to a single density $\ncrit$. Machine learning (ML) predictions of the critical density of the phase transition for each trajectory (horizontal dashed lines). (b) ML predictions of critical density $n_c$ normalized by initial density $n_0$ depending on the rescaled driving (see main text) for ${\kay > 1}$ and ${v=100 \: \rfac \gamma}$ (red) and ${v=0 \: \rfac \gamma}$ (blue), as well as for ${\kay < 1}$ (inset). The exponent $\alpha$ is tuned until all data points collapse.
}
\label{fig:collapse}
\end{figure}

In the following, we consider the system in the SOC regime, allowing Rydberg atoms to additionally decay to an inert state, effectively removing them from the system, with the rate ${b\gamma}$ (see inset of Fig.\ref{fig:collapse}a). 
% Allowing this additional loss channel, we can identify eq.~\eqref{eq:rho_scaling_dp} as the scaling of the density of ground and Rydberg state atoms around the critical driving strength \cite{helmrich2020signatures}. 
As a consequence, the system drives itself to the critical density given in MF approximation by
\begin{align}
    \label{eq:n_crit}
    n_c = \frac{\Delta \gamma}{4 \pi \Omega^2} \rfac^{-3}.
\end{align}
The SOC dynamics for different initial gas densities can be seen in Fig.~\ref{fig:collapse}a.
In the initial active phase there is a fast loss of atoms to inert states until the critical point is reached where this loss slows down substantially. To observe universal critical behavior these two time scales of atom loss must be well separated \cite{markovic2014power}.

First we consider the limit where the thermal movement of atoms occurs on a much slower timescale than the internal dynamics, rendering them effectively static. In this \textit{frozen-gas} limit, the spreading of excitations is constrained to a random Erdős–Rényi network with the average network degree $\kay$ given by \cite{brady2024griffiths}
\begin{align}
    \kay = n V_s.
\end{align}
%
%%%%%%%%%%%%%%%%%%%%%%%%%%%%%%%%%%%%%%%%%%%%%%%%%%%%%%%%%%%%%%%%%%%%%%%%%%%%%%%%%%
%%%%%%%%%%%%%%%%%%%%%%%%%%%%%%%%%%%%%%%%%%%%%%%%%%%%%%%%%%%%%%%%%%%%%%%%%%%%%%%%%%
%
\noindent At ${\kay = 1}$ a transition occurs between a non-percolating network of ground state atoms with distance $\rfac$, composed of many small disconnected clusters, and a percolating network with one large cluster on the order of the size of the system \cite{erdHos1960evolution}. For ${\kay < 1}$  this gives rise to a heterogeneous, non-universal Griffiths phase of the Rydberg facilitation, replacing the absorbing-state phase transition. Above the percolation transition, however, i.e. ${\kay > 1}$, the absorbing-state phase transition is recovered \cite{brady2024griffiths,munoz2010griffiths}. (The SOC dynamics does not change the  Erdős–Rényi  character of the network, but only leads to a reduction of $\langle k\rangle$.)

At high gas temperatures the continuous mixing of atomic positions and subsequent fast decay of spatial correlations leads to the expectation of mean field behavior regardless of $\kay$ \cite{brady2023mean}.

An unambiguous signature of universal behavior and a precise method for the classification of a given system into a certain universality class is the collapse of data obtained over a large parameter range onto specific scaling functions. Following Ref.~\cite{helmrich2020signatures}, we consider the  density of atoms in active states (i.e. in the ground and Rydberg state but not in the inert state) at the critical point, $n_c$, normalized to the initial density $n_0$ as a function of the generalized driving strength $\Omega^2 n_0^{1/\alpha}$, with $\alpha$ being tuned until all data points collapse onto a single curve. Scale invariance requires
\begin{equation}
    \frac{n_c}{n_0} = f\Bigl(\Omega^2 n_0^{1/\alpha}\Bigr) 
\end{equation}
to hold over the entire parameter range, with a scaling function $f(x)$, which can be chosen as ${f(x) = x_c^\beta (x^{\mu \beta} +x_c^{\mu \beta})^{-1 / \mu}}$ \cite{helmrich2020signatures}, where $x_c$ and $\mu$ are free parameters defining the position and sharpness of the critical point. Finally, $\beta$ corresponds to the critical exponent from eq.~\eqref{eq:rho_scaling_dp}.

For the high temperature limit the result is plotted in Fig.~\ref{fig:collapse}b (orange dots). For both ${\kay > 1}$ and ${\kay < 1}$ (inset) we receive a collapse of all data points onto a single power-law using the tuning exponent ${\alpha = 1.26(1)}$ and ${\alpha = 1.08(1)}$ respectively and we extract the critical exponents ${\beta_\text{high$\kay$} = 0.996(18)}$ and ${\beta_\text{low$\kay$} = 1.049(19)}$ respectively, which both fall in line with the expected 
%DP scaling exponent above the critical dimension, i.e. 
mean field exponent ${\beta_\mathrm{MF} = 1.00}$. Errors are calculated from the covariance matrix of the respective fit parameters.

For the frozen gas regime, on the other hand, (blue dots in Fig.~\ref{fig:collapse}b) we find no collapse of data below the percolation threshold, i.e. ${\kay < 1}$, for values of ${\alpha \in [0.5, 2.0]}$, indicating non-universal behavior which is consistent with the observation of a heterogeneous Griffiths phase \cite{brady2024griffiths}. For ${\kay > 1}$ however, the data collapses onto a single power-law for ${\alpha = 0.88(1)}$, with the slope of this power-law clearly differing from the high temperature slope. Furthermore, when using the above mentioned fit function we obtain the power-law exponent ${\beta_\mathrm{frozen} = 0.809(13)}$, very close to the expected 3D DP critical exponent ${\beta_\mathrm{DP} = 0.813}$ \cite{hinrichsen2006non}. 

% The velocity limits at which  DP and respectively MF behavior occurs is quite intuitive. If the rms velocity is less
% than the width of the facilitation shell multiplied by the facilitation rate, the effect of atomic motion can be neglected on the time scale of the facilitation and one expects DP behavior. On the other hand if the rms velocity is much larger than the facilitation distance per lifetime $\gamma^{-1}$ of the Rydberg state, an excited Rydberg atom will facilitate many atoms outside its original clusters and 
% the network structure of the frozen gas is washed out leading to MF behavior.

To unambiguously confirm DP and MF universality in the frozen-gas and high-temperature limits respectively, we also determine the critical exponent $\nu_\parallel$. To this end we 
consider the decay of excitations from the fully excited state ${\rho(t=0) = n(t=0)}$, in the number conserving case, i.e. ${b = \tau = 0}$.
For the case of weak driving, i.e. in the absorbing phase ${\kappa < \gamma}$, a pure exponential decay to ${\rho(t \to \infty) = 0}$ is expected, while for strong driving, i.e. in the active phase ${\kappa > \gamma}$, a non-zero steady state density emerges \cite{hinrichsen2006non} ${\rho(t \to \infty) > 0}$. At the critical driving strength, i.e. ${\kappa = \kappa_c = \gamma}$, there is a power-law decay in the active density of the form ${\rho \sim t^{-\delta}}$  with the exponent ${\delta = \beta / \nu_\parallel}$. For contact processes on networks, the critical driving strength is expected to be slightly larger however, i.e. ${\kappa_c \gtrsim \gamma}$ \cite{munoz2010griffiths}. For MF ${\delta = \beta = \nu_\parallel = 1}$, whereas for 3D DP universality a less steep slope with exponent ${\delta \approx 0.732}$ has been predicted, since ${\nu_\parallel^\mathrm{DP} = 1.11(1)}$, see e.g. Ref.~\cite{hinrichsen2000non}.

\begin{figure}[H]
  \centering
  \includegraphics[width=\columnwidth]{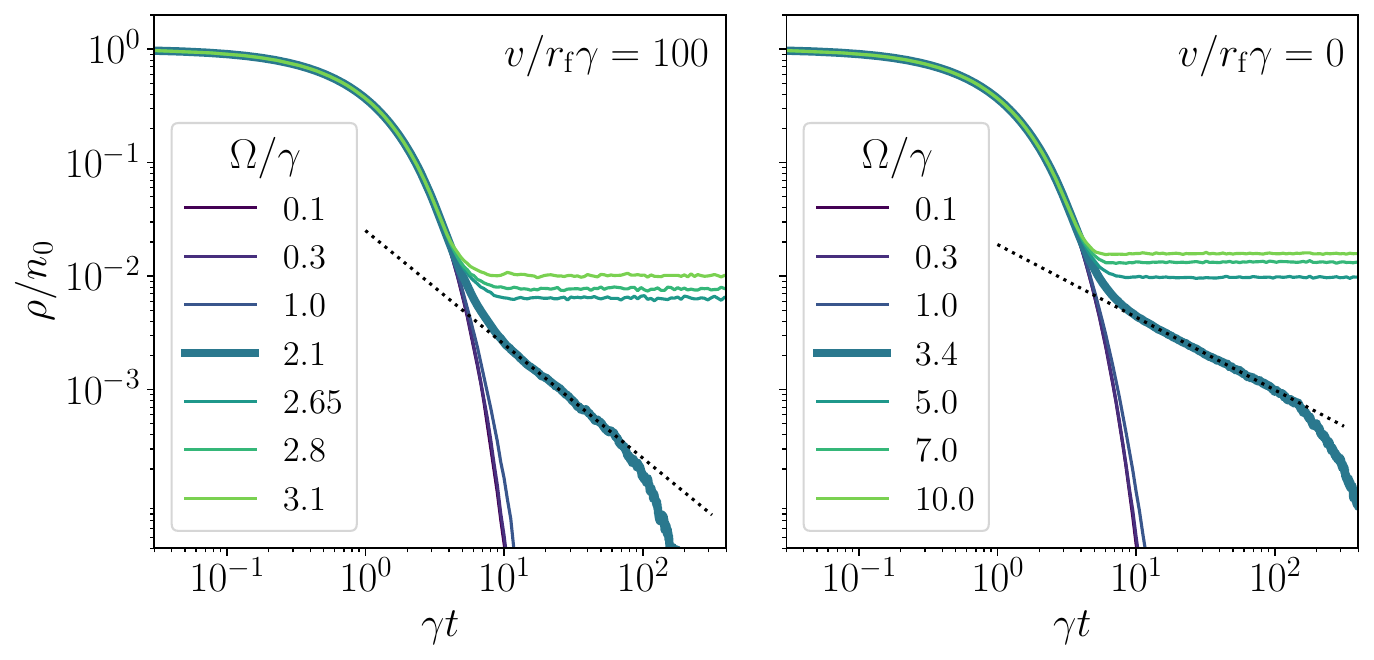}
  \caption{
        Decay of the density of excited atoms over time with all atoms initially excited ${\rho(t=0)=n_0}$ with varying driving $\Omega$ with ${n_0 \rfac^3 = 20}$ for high temperature (${v / \rfac \gamma = 100}$, left) and the frozen gas (${v / \rfac \gamma = 0}$, right). Both plots display an absorbing state phase transition with exponential decay for ${\kappa < \kappa_c \approx \gamma}$, decay to a steady state for ${\kappa > \kappa_c}$ and power-law decay for ${\kappa=\kappa_c}$. The dashed lines represent a power-law decay with exponent ${\delta=1.0(1)}$ (left) and ${\delta=0.65(10)}$ (right). Furthermore we use ${\Delta / \gamma = 1000}$ and ${b=0}$.
    }
\label{fig:delta}
\end{figure}

In Fig.~\ref{fig:delta} we see this expected behavior with an intermediate power-law for the driving strength ${\kappa(\Omega, n_0) = \kappa_c}$, which is exponentially truncated as a result of finite size effects. The behavior around the critical driving strength is very sensitive to the Rabi frequency $\Omega$, which is reflected in the rather large error margins in $\delta$. Still, we find a good agreement with the DP and MF predictions of $\delta$ and a clear difference between the values $\delta$ takes in the two limits.

%%%%%%%%%%%%%%%%%%%%%%%%%%%%%%%%%%%%%%%%%%%%%%%%%%%%%%%%%%%%%%%%%%%%%%%%%%%%%%%%%%
%%%%%%%%%%%%%%%%%%%%%%%%%%%%%%%%%%%%%%%%%%%%%%%%%%%%%%%%%%%%%%%%%%%%%%%%%%%%%%%%%%

%%%%%%%%%%%%%%%%%%%%%%%%%%%%%%%%%%%%%%%%%%%%%%%%%%%%%%%
\section{Anomalous directed percolation}
%%%%%%%%%%%%%%%%%%%%%%%%%%%%%%%%%%%%%%%%%%%%%%%%%%%%%%%

From the above discussion one would naively expect that there is a critical value of the RMS atom velocity where a phase transition  between DP and MF behavior takes place. Astonishingly we find however in Fig.~\ref{fig:beta} for  gas temperatures between the two limits (and ${\kay > 1}$) a universal collapse of data points with a \emph{monotonously changing critical exponent} $\beta$ over multiple orders of magnitude in the RMS gas velocity.

Increasing the temperature the system leaves the DP regime at rather low velocities corresponding to the (very small) width of the facilitation shell per facilitation time, i.e. ${v_- = \drfac \langle \gfac \rangle}$ (left mark in Fig.\ref{fig:beta}). This is because the number of ground-state atoms that can be facilitated by a single Rydberg atom starts to increase once this velocity is exceeded.
On the other hand the network character of ground state atoms  
gets completely washed out leading to MF behavior if the RMS velocity of a Rydberg atom is so large that it flies a distance larger than the facilitation distance 
in a facilitation time, i.e. ${v_+ = \rfac \langle \gfac \rangle}$ (right mark in Fig.\ref{fig:beta}).

In the following we show that the critical behavior with continuously varying %critical 
exponents $\beta$ in the velocity range between these two limiting values
is a signature of ADP universality, resulting from effective long-range spreading processes and heavy-tailed waiting time distributions \cite{hinrichsen2007non}.

Absorbing-state phase transitions in complex systems where excitation distances follow a Lévy flight distribution as
\begin{align}
    P(r) \sim \frac{1}{r^{d + \sigma}},
\end{align}
where $d$ is the dimension and $\sigma$ is a free parameter, no longer fulfill the constraints of the Janssen-Grassberger conjecture. Such systems however, still show universal behavior, albeit with continuously varying critical exponents depending on the value of $\sigma$
\cite{hinrichsen2006non,hinrichsen2007non}. The same is true if the distribution of time intervals between successive excitations (i.e. waiting time distribution $P(\delta t)$) is heavy-tailed. In general terms, the algebraic spatial and temporal distributions effectively reduce the upper critical dimension, and the critical exponents approach the MF values.

In the frozen gas limit each atom is confined to a cluster and has $k$ atoms in its facilitation shell, with $k$ given by a Poissonian distribution as ${P(k) = \frac{(n V_s)^k}{k!} \mathrm{e}^{-n V_s}}$. With increasing thermal velocity, the probability that an atom finds another connection outside of its original cluster increases. Since the underlying network is a random network, even small distances in real space can correspond to completely new connections, i.e. very distant jumps in the network. 

\begin{figure}[H]
  \centering
  \includegraphics[width=\columnwidth]{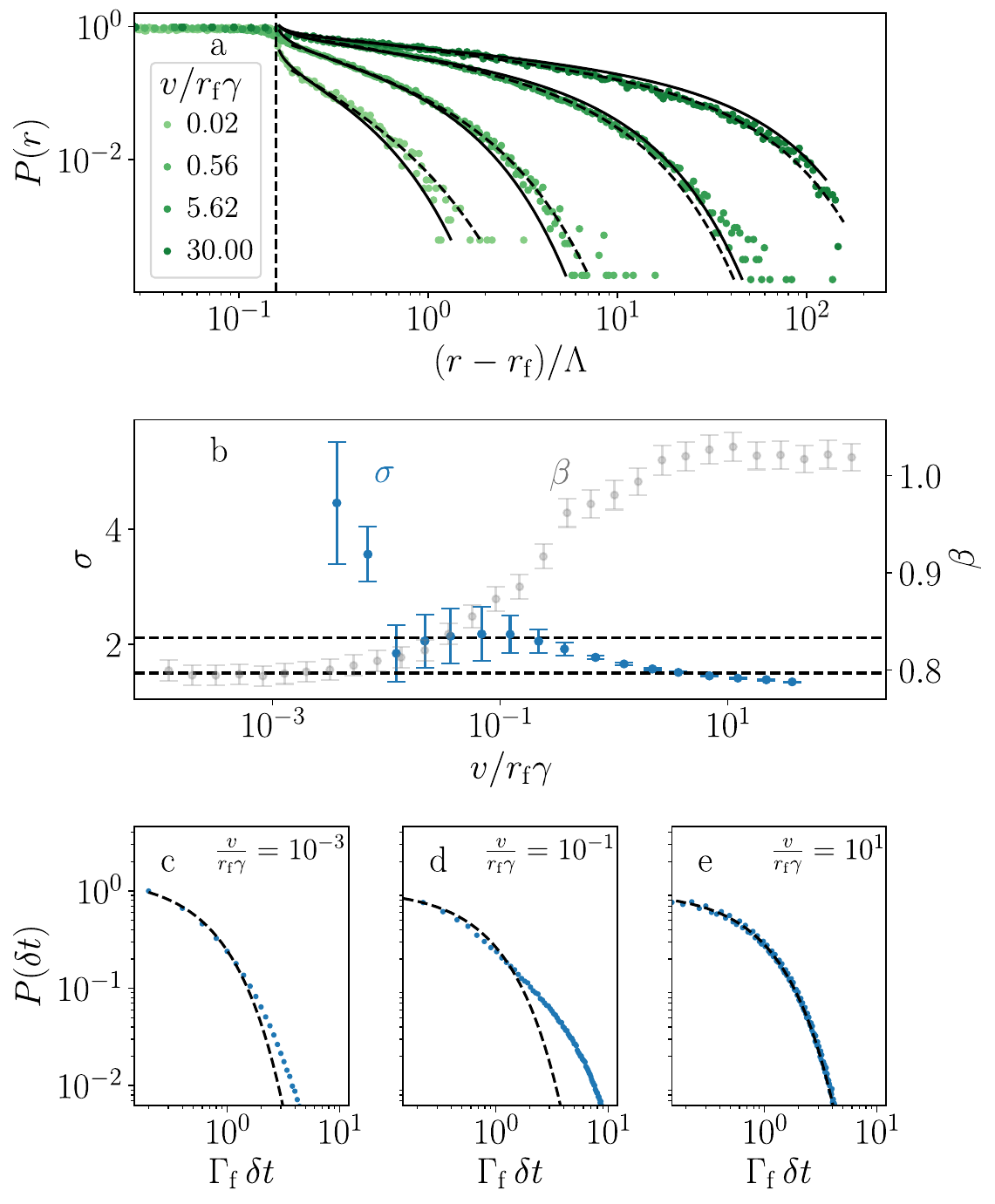}
  \caption{
        (a) Distribution of distances between excitations in units of mean free path $\Lambda$ for different RMS gas velocities $v$. Half of the width of the facilitation shell is given by the vertical gray dashed line. The black dashed lines are combined power-law/exponential fits and the black solid lines are analytical predictions given by eq.~\eqref{eq:f_of_r}. (b) Power-law exponent $\sigma$ (blue) of infection distance distribution and field theoretical limits of ADP (horizontal dashed lines, see main text), as well as critical scaling exponent $\beta$ (gray) continuously changing between MF and DP plateaus for gas velocities where $\sigma$ is within ADP bounds. And distribution of times between excitations ${P(\delta t)}$ for (c) ${v / \rfac \gamma = 10^{-3}}$, (d) ${v / \rfac \gamma = 10^{-1}}$, and (e) ${v / \rfac \gamma = 10^{1}}$. All plots use ${\Omega / \gamma = 20}$, ${\Delta / \gamma = 2000}$, ${n_0 \rfac^3 = 20}$.
    }
\label{fig:sigma}
\end{figure}

For an initially excited Rydberg atom with velocity $v$, the distribution of distances to the next facilitated atom can be seen in Fig.~\ref{fig:sigma}a. Outside of the facilitation shell (vertical black dashed line), we find that this probability decays as a power-law with an exponent $\sigma$ decreasing with increasing atom velocity. For large distances the excitation probability is exponentially truncated, with the scale given by ${v / \gfac}$.

%%%%%%%%%%%%%%%%%%%%%%%%%%%%%%%%%%%%%%%%%%%%%%%%%%%%%%%%%%%%%%%%%%%%%%%%%%%%%%%%%%
%%%%%%%%%%%%%%%%%%%%%%%%%%%%%%%%%%%%%%%%%%%%%%%%%%%%%%%%%%%%%%%%%%%%%%%%%%%%%%%%%%

In 3D systems, MF behavior occurs for ${\sigma < 1.5}$ (lower black dashed line in Fig.~\ref{fig:sigma}b), while for ${\sigma > 2.118(17)}$ regular DP behavior is expected (upper black dashed line in Fig.~\ref{fig:sigma}b) \cite{hinrichsen2006non}. In between these limits, the long range interactions are prevelant enough to disrupt DP universality, but not strong enough to suppress all correlations. Here, the system falls into the ADP universality class with its critical exponents taking on values between their MF and DP limits and which vary continuously with $\sigma$ \cite{hinrichsen2006non}.

Fitting the spatial distribution of excitation distances with an exponentially truncated power-law ${f(x) = c_1 x^{-c_2} \mathrm{e}^{-c_3 x}}$, with ${c_2 = \sigma - 1}$ (as ${P(\vec{r}) \, \mathrm{d}^3 x = P(r) 4 \pi r^2 \, \mathrm{d}r}$), we receive very good agreement between the data and the fit function (dashed lines in Fig.~\ref{fig:sigma}a). From this we can extract the power-law slope ${\sigma(v)}$ governing the flight distance distribution depending on the thermal gas velocity seen in Fig.~\ref{fig:sigma}b.

For the waiting time distribution (seen in Fig.~\ref{fig:sigma}c-e for different $v$), we find an exponential decay for the high temperature and frozen gas limits. However, for gas velocities in the interval ${v / \rfac \gamma \in [0.01, 1.0]}$  we find a deviation from a pure exponential decay in the form of a power-law for times on the order of the facilitation time $\gfac^{-1}$. This additional power-law decay in the time distribution gives rise to a regime where both spatial and temporal long-range processes are relevant (ADP II in Fig.~\ref{fig:beta}). 

For gas velocities in the interval ${v / \rfac \gamma \in [1.0, 3.0]}$, we find an exponential waiting time distribution, but a spatial power-law distribution with ${\sigma > 1.5}$, giving rise to the ADP I regime \cite{hinrichsen2000non}. Here the critical exponent of the activity scaling $\beta$ can be field theoretically approximated in perturbation theory to one-loop order, which yields \cite{hinrichsen2006non}
\begin{align}
    \beta = 1 + 2 \frac{3 - 2 \sigma}{7 \sigma}.
\end{align}
For gas velocities ${v / \rfac \gamma \gtrsim 1.0}$ we see a very good agreement between the field theoretical approximation of $\beta(\sigma(v))$ and our simulation results (black line in Fig.~\ref{fig:beta}). With decreasing velocity, i.e. entering the ADP II regime, the field theoretical predictions begin to diverge (gray dashed line in Fig.~\ref{fig:beta}) resulting from the non-exponential distribution in waiting times and the failure of the perturbation expansion.

%%%%%%%%%%%%%%%%%%%%%%%%%%%%%%%%%%%%%%%%%%%%%%%%%%%%%%%
\section{Lévy Flights}
%%%%%%%%%%%%%%%%%%%%%%%%%%%%%%%%%%%%%%%%%%%%%%%%%%%%%%%

The heavy-tailed distribution of excitation distances $P(r)$ is caused by atomic motion as we will show in the following. To this end we describe the distribution considering the distance $z$ a Rydberg atom would cover before facilitating another atom. By discretizing space into infinitesimal steps ${\delta z}$ we can write the probability that the excitation happens \textit{after} at least $J$ steps as ${P(X > J) = (1 - p_\mathrm{exc})^J}$. Here $p_\mathrm{exc}$ is the excitation probability in a given time interval ${\delta t = \delta z / v}$. As the number of atoms in the facilitation shell of a Rydberg atom in ${t + \delta t}$ are given by the Poissonian distribution, the excitation probability reads
\begin{subequations}
    \begin{align}
        p_\mathrm{exc} 
        % &= P(k=1) (1 - (1 - p_\uparrow)) \\  \nonumber
        %  &+ P(k=2) (1 - (1 - p_\uparrow)^2) + ... \\
        &= \sum_{k=0}^{\infty} P(k) (1 - (1 - p_\uparrow)^k) \\
        \label{eq:p_exc}
        &=  1 - \exp\Bigl\{-\kay \frac{\delta z}{\drfac} (1 - \mathrm{e}^{-\drfac \gfac / v})\Bigr\}.
    \end{align}
\end{subequations}
Inserting \eqref{eq:p_exc} into ${P(X \geq J)}$ and taking the limit ${\delta z \to 0}$ for fixed $z$, with ${J = z / \delta z}$ and ${J \delta z = \mathrm{const.}}$, we receive the exponential distribution as the continuum limit of the geometric distribution. The probability density function of the flight distance, for ${z \geq 0 }$, reads
\begin{align}
    f_Z(z) = \xi \mathrm{e}^{-\xi z},
\end{align}
with ${\xi = \kay / \drfac (1 - \mathrm{e}^{-\drfac \gfac / v})}$. After the Rydberg atom flies the distance $z$, an atom is facilitated in a random position around it with distance $\rfac$, i.e. in spherical coordinates with uniformly distributed random variables ${\theta \in [0, \pi)}$ and ${\varphi \in [0, 2\pi)}$, as well as a fixed ${r\equiv\rfac}$. The probability distribution of the distance from the initial position of the Rydberg to the position where the first atom is facilitated is given by %a probability transformation as
%
%\begin{subequations}
    \begin{align}
        P(r) = 
        % &\int_{\mathbb{R}^3} \mathrm{d}^3 x \; f_\Theta(\theta) f_\Phi(\varphi) f_Z(z) \delta(r - |\vec{r}|) \\
        \label{eq:f_of_r}
        &2 \pi \xi r \int_0^\pi \!\!\!\mathrm{d} \theta \, \, \frac{\mathrm{e}^{-\xi(\sqrt{\cos^2\theta + r^2 - 1} - \cos\theta})}{\sqrt{\cos^2\theta + r^2 - 1}},
    \end{align}
%\end{subequations}
%
where ${\vec{r}= \rfac \hat{e}_r(\theta, \varphi) + (0, 0, z)^T}$. In eq.~\eqref{eq:f_of_r} the distribution is given up to a numerically solvable integral and the solution can be seen as the black solid lines in Fig.~\ref{fig:sigma}a. We see an excellent agreement between the distribution given by eq.~\eqref{eq:f_of_r} and the numerical data.

%%%%%%%%%%%%%%%%%%%%%%%%%%%%%%%%%%%%%%%%%%%%%%%%%%%%%%%
\section{Methods}
%%%%%%%%%%%%%%%%%%%%%%%%%%%%%%%%%%%%%%%%%%%%%%%%%%%%%%%

%%%%%%%%%%%%%%%%%%%%%%%%%%%%%%%%%%%%%%%%%%%%%%%%%%%%%%%
\subsection{Microscopic Description of the Rydberg Gas}
%%%%%%%%%%%%%%%%%%%%%%%%%%%%%%%%%%%%%%%%%%%%%%%%%%%%%%%

We consider a three dimensional gas of $N$ atoms coupled between a ground $\ket{g}$ and a Rydberg $\ket{r}$ state with a laser with Rabi-frequency $\Omega$ and detuning $\Delta$. The unitary dynamics are described by the Hamiltonian ${\hat{H} = \sum_i \Omega \hat{\sigma}_i^x + \Delta \big(\hat{V}_i - 1 \big) \hat{\sigma}_i^{rr}}$, where $\hat{\sigma}_i^{rr}$ is the projection operator of the $i$th atom onto its Rydberg state and the Rydberg-Rydberg van-der-Waals interaction is given by the potential ${\hat{V}_i = \sum_{j<i} \frac{C_6}{r_{ij}^6}}$ with van-der-Waals coefficient $C_6$, and $r_{ij} =\vert \vec{r}_i-\vec{r}_j\vert$ being the distance between the $i$th and $j$th atom.

In addition to the unitary dynamics, we account for spontaneous decay of the Rydberg state into the ground or an additional dark $\ket{0}$ state described by the jump operators ${\hat{L}_{1,i} = \sqrt{(1 - b) \gamma} \ket{g}_{ii} \bra{r}}$ and ${\hat{L}_{2,i} = \sqrt{b \gamma} \ket{0}_{ii} \bra{r}}$ respectively. Here the parameter ${b \in [0, 1]}$ describes the portion of atoms lost from the system following decay into inert states. Finally, dephasing of the Rydberg state is accounted for by ${\hat{L}_{3,i} = \sqrt{\gamma_\perp} \ket{r}_{ii} \bra{r}}$. Typically in Rydberg many-body gases ${\gamma_\perp \gg \Omega}$, allowing classical rate equations to describe these systems to high accuracy \cite{Levi_2016}.

The evolution of the $N$-body density matrix is given by the Lindblad master equation ${\ddt \hat{\rho} = -i[\hat{H}, \hat{\rho}] + \hat{\mathcal{L}}(\hat{\rho})}$, with the Lindblad superoperator $\hat{\mathcal{L}}(\hat{\rho})$ \cite{lindblad1976generators}. After adiabatic elimination of coherences
%, i.e. ${\ddt \rho_i^{rg} = 0}$, 
a set of rate equations for the occupation probabilities in Rydberg ($p_r$) and ground states ($p_g$) of each atom can be derived. These read
\begin{subequations}
    \begin{align}
        \ddt p_r^{(i)} &= -\frac{2 \Omega^2 \gamma_\perp}{\gamma_\perp^2 + V_i^2}(p_r^{(i)} - p_g^{(i)}) - \gamma p_r^{(i)}
        \\
        \ddt p_g^{(i)} &= +\frac{2 \Omega^2 \gamma_\perp}{\gamma_\perp^2 + V_i^2}(p_r^{(i)} - p_g^{(i)}) + \gamma p_r^{(i)},
\end{align}
\end{subequations}
with ${V_i = \Delta \big(-1 + \sum_{j \in \Sigma} \frac{\rfac^6}{r_{ij}^6} \big)}$, where $\Sigma$ is the set of all other atoms in the Rydberg state.

For all simulations we initiate random positions in a 3D box with length ${L = 7 \rfac}$ and periodic boundary conditions. Atom velocities are sampled from a Maxwell-Boltzmann distribution, i.e. Gaussian in each direction, with the temperature given by the RMS velocity $v$. Furthermore, we use ${\gamma_\perp = 20}$ and a fixed time step Monte-Carlo algorithm with time step ${\gamma \, dt = 0.0025}$.

%%%%%%%%%%%%%%%%%%%%%%%%%%%%%%%%%%%%%%%%%%%%%%%%%%%%%%%
\subsection{Machine Learning the Critical Density}
%%%%%%%%%%%%%%%%%%%%%%%%%%%%%%%%%%%%%%%%%%%%%%%%%%%%%%%

Upon the system reaching its critical density, the total density continues to decay as a result of a finite separation of time-scales in the system (see Fig.~\ref{fig:collapse}a), which makes the unambiguous determination of the true critical point difficult. Therefore, we developed a machine learning algorithm which learns the critical density based on the time dependent total density of the gas. For the case of high gas temperatures, specifically an average thermal velocity ${v_\mathrm{th}=100 \: \rfac \gamma}$, the dynamics of the system are accurately described by mean field equations \cite{brady2023mean}. For the mean field limit the critical density $n_c$ is given by eq.~\eqref{eq:n_crit}.

We then train the algorithm by passing it a sub-sampled vector $X$, consisting of 100 density points from $n(t)$, which are equally spaced in time. The times are spaced in the interval ${\gamma t \in [0, 2000]}$ and have a spacing of ${\gamma \, \delta t = 20}$. All training trajectories use ${v_\mathrm{th}=100 \rfac \gamma}$, i.e. are in the mean field limit, and are the average of 50 Monte-Carlo runs. As all training data is in the percolating limit, the system consists of roughly $10^4$ atoms, resulting in a fast convergence of Monte-Carlo simulations.

In total, the algorithm is trained using 5866 trajectories, each with a unique combination of $\Omega$ and $n_0$. The algorithm outputs ${Y \in [0, 1]}$ and predicts ${\bar Y \equiv n_c(\Omega) / n_0}$ for ${n_c < n_0}$ and ${\bar Y \equiv 1}$ for ${n_c \geq n_0}$, where we found a modified Huber-Loss function \cite{huber1964} to give the most accurate predictions. This is defined as 
\begin{align}
    L_\Delta(Y - \bar Y) = 
    \begin{cases}
        100 \times \frac{1}{2} \frac{|Y - \bar Y|^2}{\bar Y}, \;\; |Y - \bar Y| \leq \Delta
        \\
        100 \times \Delta \frac{|Y - \bar Y|}{\bar Y}, \;\; |Y - \bar Y| > \Delta
    \end{cases}.
\end{align}
Using ${\Delta = 1}$ yields an average percentage error of 4~\%. We then apply the algorithm to determine ${\ncrit / n_0}$ for arbitrary $v_\mathrm{th}$.

%%%%%%%%%%%%%%%%%%%%%%%%%%%%%%%%%%%%%%%%%%%%%%%%%%%%%%%
\subsection{Critical Density Scaling}
%%%%%%%%%%%%%%%%%%%%%%%%%%%%%%%%%%%%%%%%%%%%%%%%%%%%%%%

It has been shown for this system that the critical gas density $n_c$ separating the absorbing from the active phase exhibits scale-invariant behavior with respect to the driving strength and the initial density of the SOC dynamics. Applying the scaling ansatz ${n_c = n_0 f(\Omega^2 n_0^{1 / \alpha})}$ and tuning the exponent $\alpha$ until all curves collapse onto a single power-law, following \cite{helmrich2020signatures}, the critical exponent $\beta$ of the directed percolation universality class can be extracted.

In order to find the exponent $\alpha$ where all curves collapse, we choose exponents in the interval ${\alpha \in [0.5, 2.0]}$. For each exponent we fit the data using the heuristic fit function defined in the main text and defining a loss function as the mean squared error between the data points scaled with $\alpha$ and a fit using ${f(x) = x_c^\beta (x^{\mu \beta} +x_c^{\mu \beta})^{-1 / \mu}}$ and determine the mean squared error between the data points and $f(x)$. For all percolating datasets, as well as for the non-percolating high temperature dataset, we find a convex loss function with a well defined minimum where the data points collapse. Additionally, we find no universal collapse of data for the low temperature, non-percolating case (see Fig.~\ref{fig:collapse}).

\subsection*{Acknowledgement}

Financial support from the DFG through SFB TR 185, project number
277625399, is gratefully acknowledged. 

The simulations were [partly] executed on the high performance cluster “Elwetritsch” at the RPTU Kaiserslautern which is part of the “Alliance of High Performance Computing Rheinland-Pfalz” (AHRP). We kindly acknowledge the support of the RHRZ.

\subsection*{Authors contributions}

MF conceived and supervised the project.
Monte-Carlo numerical simulations were performed by DB with support from SO and JO. The machine learning algorithm was conceived by DB with support from SO and JO and implemented together by DB and SO. All authors discussed the results and developed a physical understanding describing the numerical findings. DB and MF wrote the initial version of the manuscript with support by SO.

% MC numerical simulations by D.B. with support of S.O.  Machine Learning algorithm conceived by S.O. and implemented by S.O. together with D.B. Project conceived and supervised by M.F. J.O. support in ML algorithm. All authors discussed results, developed physical picture describing numerical and experimental findings. D.B. and M.F. wrote the initial version of the mansucript with support by S.O. 

\bibliography{references}

\end{document}